\documentclass[fleqn,usenatbib]{mnras}

\usepackage{newtxtext,newtxmath}
\usepackage{graphics}
\usepackage{epsfig}	
\usepackage{amsmath}	

\usepackage{amssymb}	
\usepackage[T1]{fontenc}

\DeclareRobustCommand{\VAN}[3]{#2}
\let\VANthebibliography\thebibliography
\def\thebibliography{\DeclareRobustCommand{\VAN}[3]{##3}\VANthebibliography}

\title[Mapping FRB]{Classifying FRB spectrograms using nonlinear dimensionality reduction techniques}

\author[X. Yang et al.]{
X. Yang$^{1,2}$, S.-B. Zhang$^{1}\thanks{sbzhang@pmo.ac.cn (S-BZ)}$, J.-S. Wang$^{3}\thanks{jiesh.wang@gmail.com (J-SW)}$, X.-F. Wu$^{1,2}\thanks{xfwu@pmo.ac.cn (X-FW)}$
\\
$^{1}$ Purple Mountain Observatory, Chinese Academy of Sciences, Nanjing 210023, China\\
$^{2}$ School of Astronomy and Space Sciences, University of Science and Technology of China, Hefei 230026, China\\
$^{3}$ Max-Planck-Institut f\"ur Kernphysik, Saupfercheckweg 1, D-69117 Heidelberg, Germany\\
}

\pubyear{2023}

\begin{document}
\label{firstpage}
\pagerange{\pageref{firstpage}--\pageref{lastpage}}
\maketitle

\begin{abstract}
Fast radio bursts (FRBs) are mysterious astronomical phenomena, and it is still uncertain whether they consist of multiple types. In this study we use two nonlinear dimensionality reduction algorithms - Uniform Manifold Approximation and Projection (UMAP) and t-distributed stochastic neighbour embedding (t-SNE) - to differentiate repeaters from apparently non-repeaters in FRBs. Based on the first Canadian Hydrogen Intensity Mapping Experiment (CHIME) FRB catalogue, these two methods are applied to standardized parameter data and image data from a sample of 594 sub-bursts and 535 FRBs, respectively. Both methods are able to differentiate repeaters from apparently non-repeaters. The UMAP algorithm using image data produces more accurate results and is a more model-independent method. Our result shows that in general repeater clusters tend to be narrowband, which implies a difference in burst morphology between repeaters and apparently non-repeaters. We also compared our UMAP predictions with the CHIME/FRB discovery of 6 new repeaters, the performance was generally good except for one outlier. Finally, we highlight the need for a larger and more complete sample of FRBs.
\end{abstract}



\section{Introduction}
Fast Radio Bursts (FRBs) are bright radio pulses with a short duration of typical milliseconds. 
Since the first FRB was discovered in Parkes archival observation~\citep{Lorimer:2007qn}, more than 600 have been detected by telescopes worldwide~\footnote{The FRB online Catalogue is available from \url{www.wis-tns.org}}. 
In general, FRBs have dispersion measures (DMs) that exceed what can be attributed to our Galaxy~\citep{Yao2017}, and they are considered to be extragalactic. However, one FRB (FRB~200428) has been shown to be originated from a Galactic magnetar~\citep{STARE2,chime200428}.
Furthermore, approximately 20 FRB sources have been localized to their host galaxies to date~\citep[e.g.][]{Chatterjee17,Bannister19,DSA10,Prochaska19,askap_frb190711,Marcote20}.

While the majority of FRBs are seen as one-off events, nearly 30 repeaters have been identified~\citep{Spitler:2016dmz,CHIME19_2r,CHIME19_8r,Kumar19,CHIME20_9r,FRB20200120E_in_globular,FRB20190520B,FRB20201124A}, leading to the division of FRBs into two classes based on their apparent repeatability.  As the population of FRBs has rapidly increased and their properties have been studied in more detail, a better understanding of their classification has emerged.
For example, \cite{volume_rate} concluded that most FRB events must originate from repeaters based on the evidence of their volumetric occurrence rate. Additionally,  \cite{energy_distribution} found that the Canadian Hydrogen Intensity Mapping Experiment (CHIME) repeater sample is consistent with the energy dependence of the apparently non-repeating Australian Square Kilometre Array Pathfinder~(ASKAP) sample, suggesting that they may come from the same population.

Alternatively, with the rapid growth of the FRB samples, possible sub-populations are gradually arising among the repeating or apparently non-repeating FRBs.
%
Most of the repeaters are highly linearly polarized~\citep[][]{121102_polar,linear_polarisation}, but FRB180301 shows a diversity of polarimetric properties~\citep[][]{luo_polar}. Only two out of the nearly 30 repeaters show a periodicity~\citep[][]{121102_period,180916_period}. The repetition rates of some repeaters are reported much lower than those notable repeating FRBs like FRB 20121102A and FRB 20201124A~\citep[][]{repeat_rate,fast_121102,fast_20201124a}.
\cite{frbtype} apply Kolmogorov-Smirnov (KS) test in CHIME/FRB catalogue, and introduce four new sub-types of repeaters and apparently non-repeaters with respect to the cosmic star formation history (SFH). However, the final confirmation of any sub-populations for the repeating or apparently non-repeating FRBs still needs more completed samples and higher sensitivity observations.

The CHIME/FRB Collaboration published its first FRB catalogue in 2021~\citep[][]{chimecatalog}, which is the first large-sample homogeneous FRB catalogue. The catalogue includes 48 kinds of observation parameters and time-frequency images of 536 bursts, including 474 apparently non-repeating bursts and 62 bursts from 18 repeating sources.
To classify repeaters and apparently non-repeaters in the catalogue, \cite{umap_frb} utilized an unsupervised machine learning algorithm based on the Uniform Manifold Approximation and Projection (UMAP). 
Their classification provides a repeating FRB completeness of 95\% and identifies 188 FRB repeater source candidates. This research enlightens us about the importance of dimensionality reduction techniques for accurately classifying FRBs.

Dimensionality reduction techniques are useful for visualizing high-dimensional data.
Principal component analysis (PCA) is a well-established linear technique for dimensionality reduction that involves combining the original variables to obtain principal components.
However, nonlinear techniques have been increasingly popular in recent years and are better suited for addressing overcrowding problems in data feature extraction.
The t-Distributed Stochastic Neighbor Embedding (t-SNE) was the most commonly used nonlinear technique in single-cell analysis before 2018~\citep{tsne}, but UMAP has since become more popular among scientists due to its speed and ability to retain both local and global structures of data~\citep{umap0}. 
%
In the field of biology, UMAP is known for providing the fastest run times, highest reproducibility, and most meaningful organization for cell clusters~\citep{umap5}. In addition, proximity in low dimensional UMAP space identifies groups of genes and finds novel protein interactions~\citep{umap10}. 
%
UMAP has also been proven efficient in clustering astronomical data, including separating H$\alpha$-emission spectra, randomly selecting spectra without H$\alpha$-emission~\citep{umap_LAMOST}, and clustering auroral images~\citep{umap_Auroral}. 
In these researches, UMAP has been applied to various input datasets, such as mass cytometry, single-Cell RNA Sequencing, expression profile data, image and H$\alpha$-emission spectra. 

In this paper, we utilize the datasets from the CHIME/FRB Catalogue to investigate the morphology of FRBs and their classification using nonlinear dimensionality reduction techniques. We apply these techniques to both the time-frequency images and standardized high-dimensional observation parameters of FRBs in order to identify potential sub-populations, and further explore the impact of morphology on FRB classification.
In Section~\ref{sec:data}, we provide a description of the data and algorithms used in this study. The results of our clustering analysis are presented in Section~\ref{sec:result}, while Section~\ref{sec:Dis} includes a discussion of our findings.

\section{Implementation}
\label{sec:data}
\subsection{Data selection}
The data used in this study were obtained from the first CHIME/FRB catalogue~\citep{chimecatalog}, which includes 536 bursts or 600 sub-bursts detected in the observations from July 25, 2018 to July 1, 2019. 
In instances where multiple peaks appeared in the light curves, the FRBs were divided into sub-bursts, resulting in a total of 600 sub-bursts. Of these, 506 sub-bursts were identified in 474 apparently non-repeating FRBs, while 94 sub-bursts were detected in 18 repeating FRBs.

To ensure comparability, the input data used for dimensionality reduction were divided into two types: images with the same scale, and parameters converted to the same scale. We followed the recommendation in the UMAP documentation to convert each feature to the same scale.

\begin{figure*}
	\includegraphics[width=0.98\textwidth]{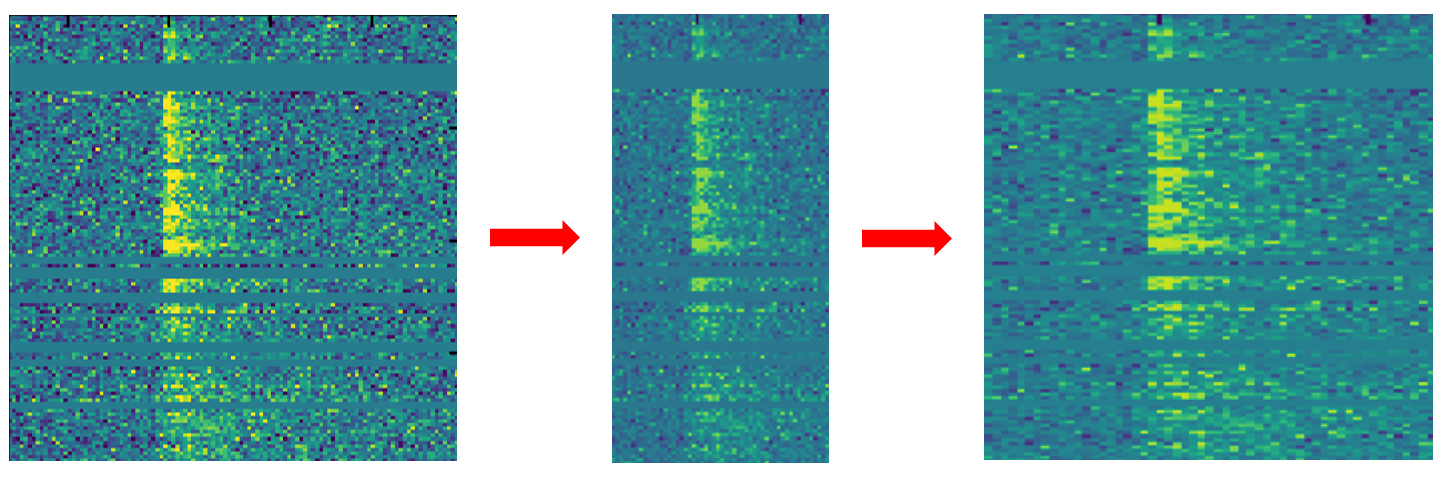}
    \caption{A sample of the extraction process of image data. The left panel shows the \emph{\sc cfod} standard output figure, the middle panel shows the extracted burst area and the right panel shows the final input data for Resnet.}
    \label{figure:1}
\end{figure*}

%
(1) 
The first type of data used in this study were images containing the dynamic spectrum of the dedispersed burst with time and frequency averaged.
We selected 535 FRB images, which included 599 FRB sub-bursts detected by the CHIME/FRB Project~\footnote{The image of FRB20190626A is not provided by the CHIME/FRB catalogue website.}. These images were processed using \emph{\sc cfod} python package, and an example of the standard output is presented in the left panel of Figure~\ref{figure:1}.  

%
(2) The second type of data consisted of the parameters of the sub-bursts, which included ten observed properties and three inferred properties.
To facilitate comparison with the work of~\cite{umap_frb}, we selected the same parameters, namely: boxcar Width, width of sub-burst, flux, fluence, scattering time, spectral index, spectral running, highest frequency, lowest frequency, peak frequency, redshift, radio energy, and rest-frame intrinsic duration. Since flux measurements were not available for six sub-bursts, this study used 13 parameters for the 594 sub-bursts.

\begin{figure*}
	\includegraphics[width=0.98\textwidth]{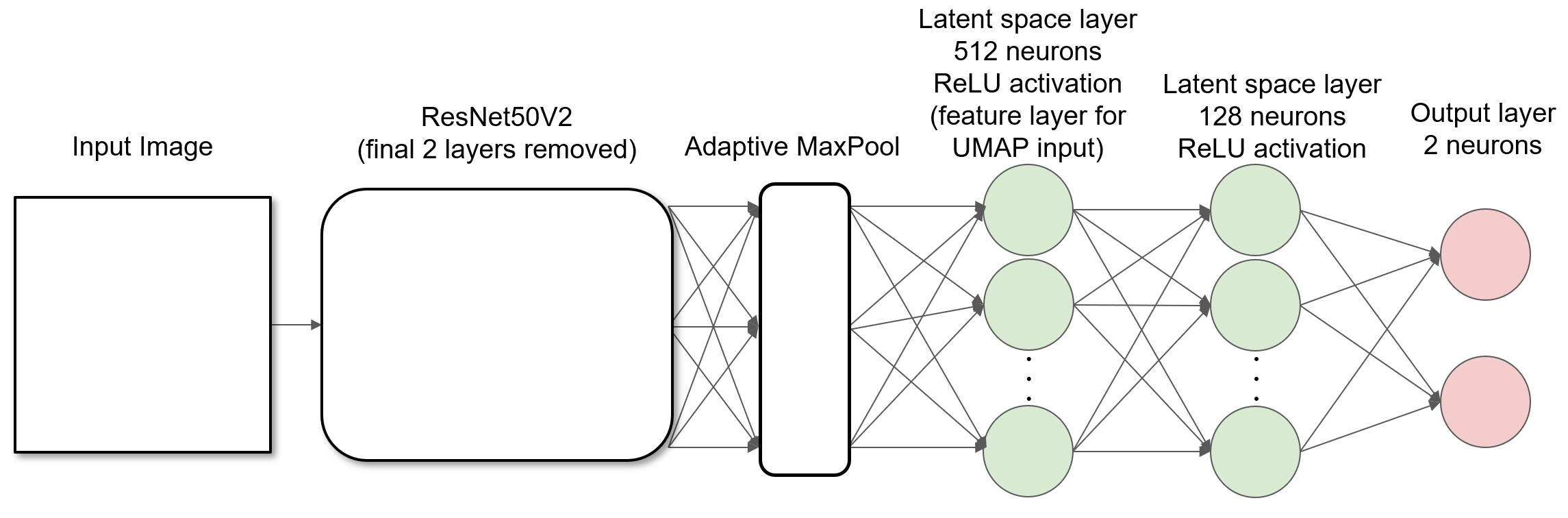}
    \caption{The $256\times256$ FRB image is utilized as the input image, and subsequently passed into the pre-trained ResNet50V2 model. The final two layers of the pretrained network have been substituted with an adaptive max pooling layer, which is followed by two fully connected layer with 512 and 128 neurons, respectively. These two layers are activated by ReLU. The final output is a fully connected layer with 2 neurons.}
    \label{figure:resnet}
\end{figure*}

\subsection{Data preparation}

Each of the 535 image data generated by \emph{\sc cfod} contains $911\times870 = 792,570$ pixels.
To avoid under-fitting,  
we extracted the burst area from the image, as shown in the middle panel of Figure~\ref{figure:1}. 
In this process, we extract multiple sub-bursts with full width at half maximum (FWHM) that overlap each other as a single image. 
Seven of the 535 bursts have two distinct sub-bursts, resulting in a total of 542 images.
To augment the data, we varied the width of each extracted image randomly along the time axis. As shown in Figure~\ref{figure:1}, the cropped width of the middle image varied each time. For each FRB, we generated five images with random widths between two to four times the FWHM of the FRB, resulting in 2,710 images.  
We then employed a deep residual network (ResNet) model to extract features from each image and obtain a 512-dimensional feature vector that represents a compressed version of the original image. The 2,710 images were resized to a dimension of 256x256 pixels and fed into the ResNet model. 
The process we followed is similar to the methodology proposed by \cite{res_umap}. Given the limited number of samples available for training, we opted for ResNet50V2, which contains 2.4 times fewer parameters than ResNet152V2.
To fine-tune the pre-trained weight of ResNet50V2, we used our labeled and augmented dataset comprising 1,200 images, including 600 labeled repeating FRBs and 600 labeled non-repeating FRBs. 
As illustrated in Figure~\ref{figure:resnet}, we eliminated the final two layers and added an adaptive max pooling layer, along with two fully connected layers containing 512 and 128 neurons, respectively, activated by rectified linear units (ReLU). 
Considering the presence of hidden repeaters in apparently non-repeating FRBs, we did not expect the ResNet50V2 model to precisely predict the labels. 
Following 100 training epochs, our ResNet50V2 implementation achieved a best accuracy of 83.7\% on the test set.

The selected thirteen parameters contain six kinds of units, three of which, including spectral index, spectral running, and redshift are dimensionless.   
To prevent the possible effect of parameter units in clustering results, instead of using the original values of these parameters, we apply z-score standardization~(i.e., the number of standard deviations from the mean) to them before training. 
This step is defaulted by the documentation of UMAP~\footnote{More detailed information can be found in \url{umap-learn.readthedocs.io}\label{web1}}, and this is the main difference between our input parameters and \cite{umap_frb}.

\subsection{UMAP and t-SNE model configuration}

We implemented python packages of the UMAP and t-SNE algorithms in this study. UMAP is known to preserve global data structure well, while t-SNE is better at presenting local data structure~\citep{umap5}. Combining UMAP and t-SNE allows us to gain a comprehensive understanding of the data from different perspectives.

(1) To optimize the clustering results, we tuned three parameters for UMAP~\textsuperscript{\ref {web1}}: \emph{\sc n\_neighbors}, \emph{\sc min\_dist}, \emph{\sc n\_components}. 
\emph{\sc n\_neighbors} represents the size of the local neighbourhood and affects the balance between local and global structure in the data. Larger values of \emph{\sc n\_neighbors} will make UMAP focus on a broader structure of the data but may lose more detailed information. 
\emph{\sc Min\_dist} controls the minimum distance between points in the low dimension and helps to preserve the broad topological structure. 
\emph{\sc N\_components} allows the user to choose the dimensionality of the reduced dimension, with two or three dimensions being common in data visualization. 
We performed a grid search for these parameters, \emph{\sc n\_neighbors} range from 2 to 100, \emph{\sc min\_dist} range from 0.01 to 0.99. Finally 
\emph{\sc n\_neighbors=15}, \emph{\sc min\_dist=0.05} and \emph{\sc n\_components=2} are determined for the Resnet image feature dataset. \emph{\sc n\_neighbors=8}, \emph{\sc min\_dist=0.1} and \emph{\sc n\_components=2} were determined for the standardized parameter data.


(2) For t-SNE, we adjusted five parameters to optimize clustering~\footnote{More detailed information can be found in \url{scikit-learn.org}}: \emph{\sc perplexity}, \emph{\sc early\_exaggeration}, \emph{\sc learning\_rate}, \emph{\sc n\_iter}, and \emph{\sc n\_components}. 
\emph{\sc Perplexity} is similar to \emph{\sc n\_neighbors} in UMAP, affects the nearest neighbours in the algorithm and is recommended to be set to larger values for larger datasets. 
\emph{\sc Early\_exaggeration} changes the tightness of natural clusters in the original embedded space and the space between natural clusters.
\emph{\sc Learning\_rate} controls the speed at which t-SNE updates its parameters and can affect the amount of input information preserved in the embedded data.
\emph{\sc N\_iter} represents the number of iteration steps for the optimization. 
As with UMAP, \emph{\sc n\_components} determines the dimensionality of the reduced dimension. 
We performed a grid search for these parameters, \emph{\sc perplexity} range from 5 to 100, \emph{\sc early\_exaggeration} range from 10 to 80, \emph{\sc learning\_rate} range from 10 to 1000. Finally we selected \emph{\sc perplexity=11}, \emph{\sc early\_exaggeration=46}, \emph{\sc learning\_rate=510}, \emph{\sc n\_iter=5000}, and \emph{\sc n\_components=2} for the Resnet image feature dataset, and \emph{\sc perplexity=15}, \emph{\sc early\_exaggeration=12}, \emph{\sc learning\_rate=200}, \emph{\sc n\_iter=5000}, and \emph{\sc n\_components=2} for the standardized parameter data. While parameters within a reasonable range can yield acceptable results, the grid search helped us choose the best parameters for our study.  


\section{Results and Discussion}
\label{sec:result}

\subsection{UMAP and t-SNE training result}

\begin{figure*}
    \begin{center}
    \includegraphics[width=0.95\textwidth]{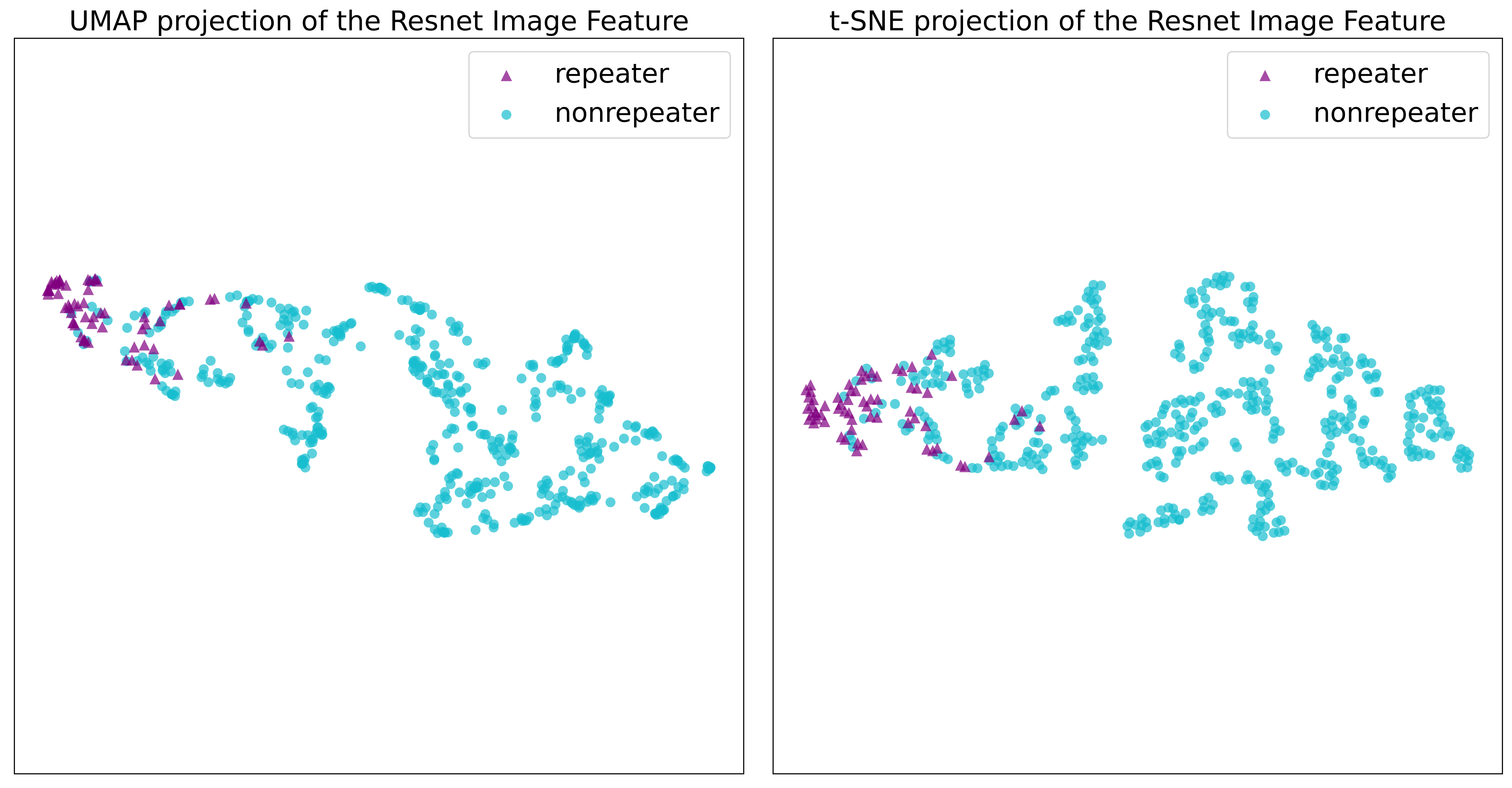}
    \caption{The UMAP and t-SNE embeddings of 542 FRB
Resnet image features. The pink triangles represent repeaters, and the cyan dots represent non-repeaters.}
    \label{figure:resimg}
    \end{center}
\end{figure*}

In Figure~\ref{figure:resimg}, the results of applying UMAP and t-SNE to the ResNet image features of 542 FRBs are presented. The UMAP and t-SNE projections clearly show that repeaters were tightly clustered together in a specific region, while a noticeable gap was present between the mixture and pure non-repeaters. These ResNet image features exhibited excellent performance in distinguishing between repeaters and non-repeaters, indicating their potential as an effective tool for identifying repeating FRBs.

\begin{figure*}
    \begin{center}
    \includegraphics[width=0.95\textwidth]{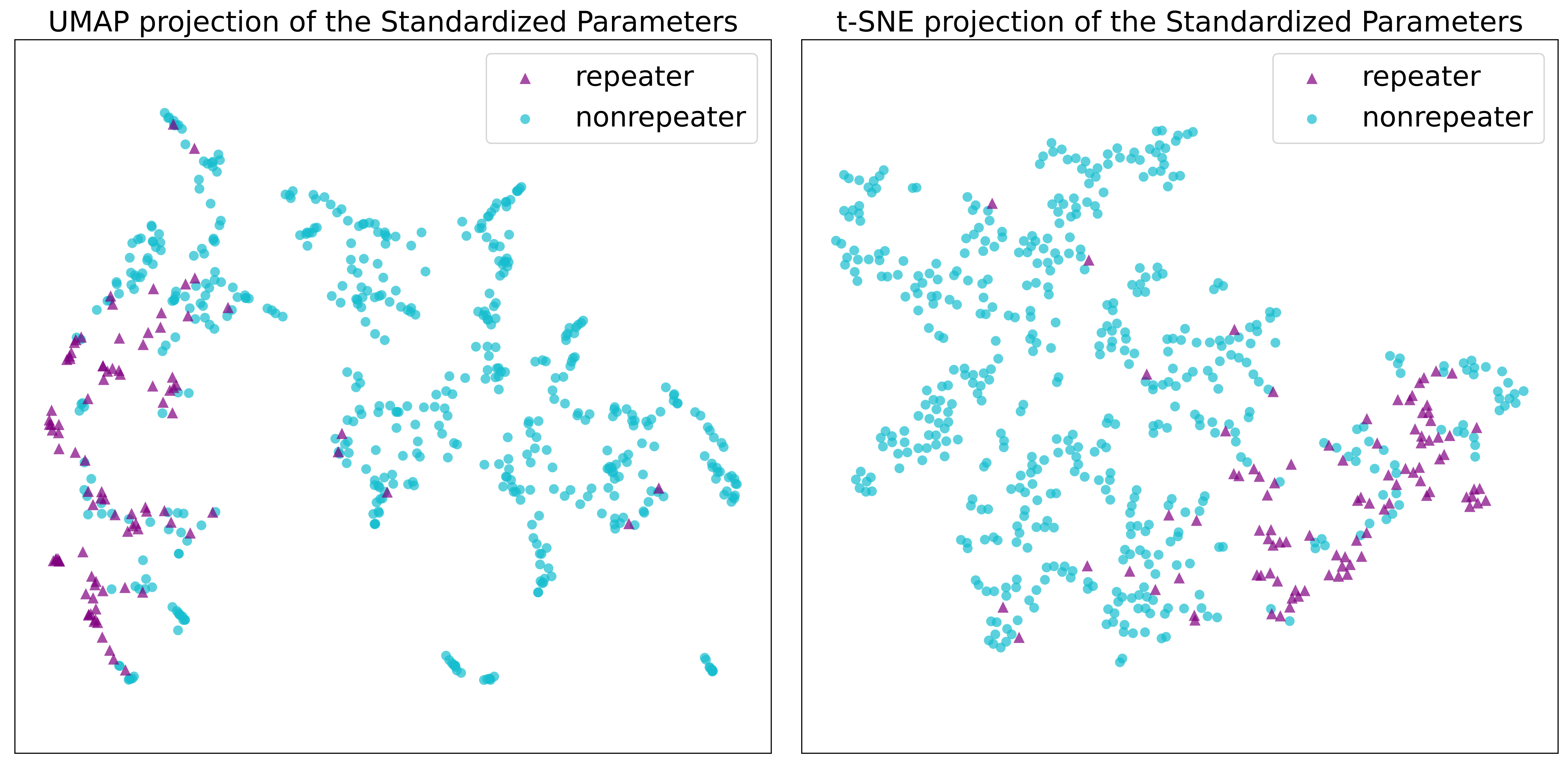}
    \caption{The UMAP and t-SNE embeddings of 594 FRB
standardized parameters. The pink triangles represent repeaters, and the cyan dots represent non-repeaters.} 
    \label{figure:3}
    \end{center}
\end{figure*}

For comparison with different input data, the FRB samples containing thirteen standardized parameters are also reduced and clustered. Figure \ref{figure:3} shows the UMAP and t-SNE embeddings of 594 FRB events. 
%
%
Although we used the same thirteen parameters as~\cite{umap_frb}, our results are very different from their separated distributions.  
Moreover, our results for the parameter data are similar to those for the image data. Most of the repeaters are spread in a quarter of the two-dimensional embedding plane, while the apparently non-repeaters are randomly scattered on the whole plane. Both UMAP and t-SNE successfully extract the difference between repeaters and apparently non-repeaters in the outline. 



\subsection{Classifying FRBs by the image and parameter data}
\begin{figure*}
    \begin{center}
    \includegraphics[width=0.95\textwidth]{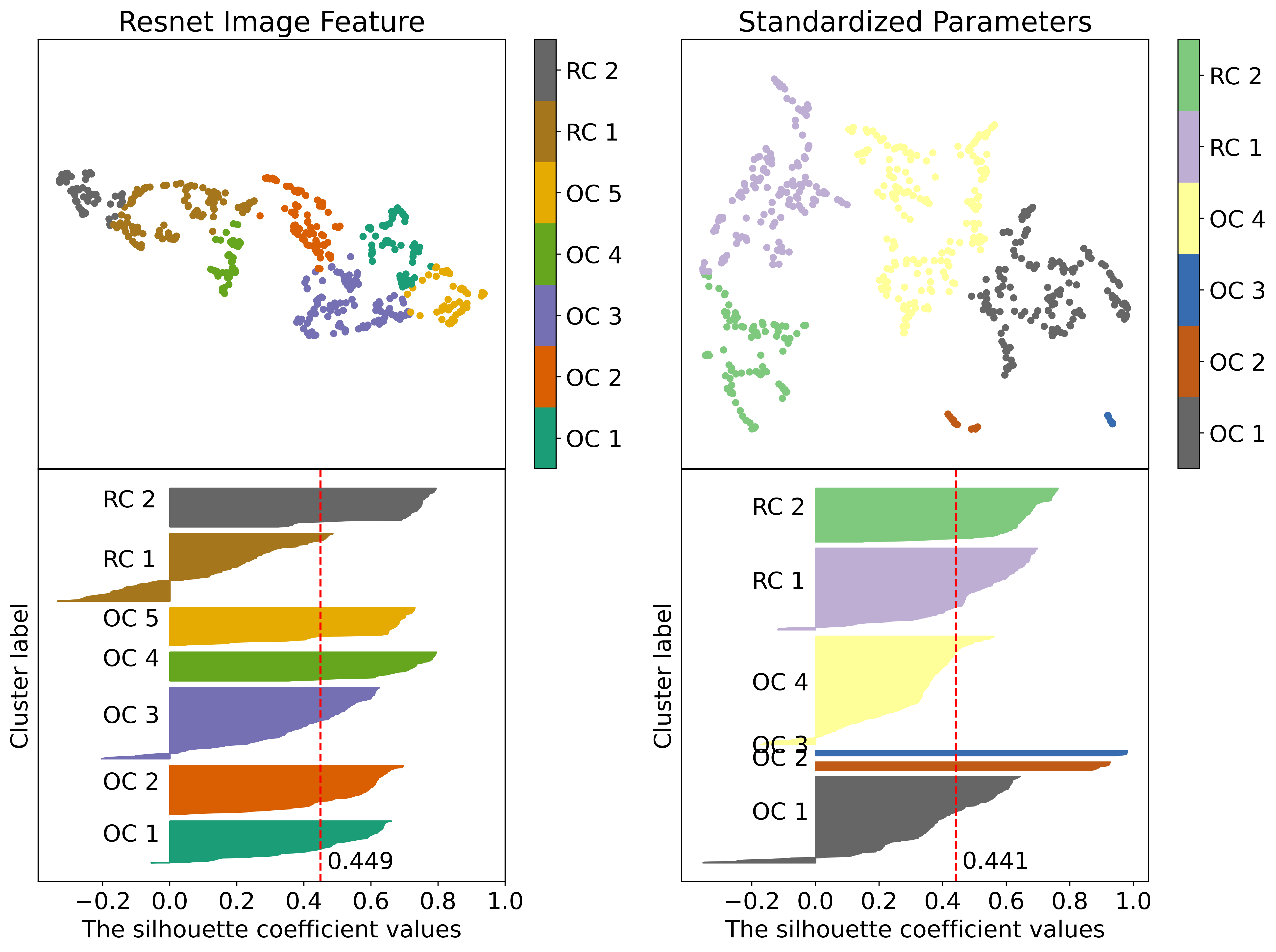}
    \caption{The spectral clustering result of UMAP embedding. The label OC means One-off cluster and RC means Repeater cluster in table~\ref{table:rate2} and table~\ref{table:rate3}. The second row are the corresponding silhouette coefficient values. The red dotted lines in the second row represent the average silhouette coefficient values of the clusters.}
    \label{figure:sp_cluster}
    \end{center}
\end{figure*}

Spectral clustering is one of the most popular modern clustering algorithms, and clustering results obtained using this approach often outperform the traditional approaches like k-means~\citep{spectral_clustering}. 
It works by using the eigenvectors of a similarity matrix to partition the data into clusters. In this algorithm, the data points are treated as nodes in a graph, and clustering is approached as a graph partitioning problem.
Using the spectral clustering algorithm, we clustered the UAMP two-dimensional embeddings~\footnote{Although t-SNE also seems to find the different distribution between two kinds of FRBs, the structure of its embedding makes the spectral clustering result unsatisfactory.}. Based on the percentage of repeaters in the clusters, we named them either repeater clusters or one-off clusters. The apparently non-repeaters in the repeater cluster are labeled as repeater candidates. Figure \ref{figure:sp_cluster} shows the spectral clustering results of the image and parameter data. To evaluate the performance of our model, we defined:
\begin{equation}
\begin{split}
    {\rm Recall}=\frac{T_{\rm p}}{T_{\rm p}+F_{\rm n}},
	\label{eq:slim}
\end{split}
\end{equation}
where $T_{\rm p}$ is true positive, indicating the number of repeaters correctly retrieved by the model in the repeater clusters,
and $F_{\rm n}$ is false negative, indicating the number of repeaters incorrectly retrieved by the model in the apparently non-repeater clusters.

To optimize the clustering results, a grid search was performed with the number of clusters ranging from 2 to 10. During this search, both the recall and silhouette coefficient values were considered to evaluate the effectiveness of the clustering. The Silhouette Coefficient is a measure of how well-defined the clusters are, and it ranges between -1 and 1, with values closer to 1 indicating more coherent clusters. A silhouette coefficient greater than 0 indicates an acceptable result. The second row of Figure \ref{figure:sp_cluster} displays the silhouette coefficient values of the clusters, with the red dotted line indicating the average silhouette coefficient values, which were greater than 0.4.


%
As shown in the middle panel of Figure \ref{figure:sp_cluster}, the UAMP embedding for Resnet image feature data is clustered by the spectral clustering algorithm. The recall of 542 samples is 100.0\%. Table \ref{table:rate2} lists the detailed content of 7 clusters, including 2 repeater clusters and 5 one-off clusters.

As shown in the right panel of Figure \ref{figure:sp_cluster}, the UAMP embedding for standardized parameters data is also clustered by the spectral clustering algorithm. The recall of 594 samples is 94.7\%, with 5 repeater samples falling out of repeater clusters.
%
Table \ref{table:rate3} lists the detailed content of 6 clusters, including 2 repeater clusters and 4 one-off clusters. From all 500 apparently non-repeater sub-bursts, 145 repeater candidates were identified~\footnote{These sub-burst candidates refer to 134 apparently non-repeating sources}.  
Assuming they are real repeaters, the FRB repeating rate is estimated to be 30.9\%, lower than the predicted rate of 41.9\% by ~\cite{umap_frb}.   

\begin{table*}

\caption{The information of the spectral clustering result of UMAP Resnet image feature dataset embedding. There are two repeater clusters, the apparently non-repeaters in repeater clusters are classified as repeater candidates.\label{table:rate2}}
\renewcommand\arraystretch{1.5}    
\begin{center}
\begin{tabular}{c|c|c|c|c}
\hline

\multicolumn{1}{c|}{ }  &     Total number        &     Repeater number  	 &     	Repeater candidate number  & Apparently non-repeater number        \\\hline

One-off cluster 1    &   68  &   0   &   0     & 68     \\\hline
One-off cluster 2    &   79  &   0   &   0     & 79     \\\hline
One-off cluster 3    &   115  &   0   &   0     & 115     \\\hline
One-off cluster 4    &   47  &   0   &   0     & 47     \\\hline
One-off cluster 5    &   61  &   0   &   0     & 61     \\\hline
Repeater cluster 1    &   109  &   16   &    93    & 0   \\\hline 
Repeater cluster 2    &   63  &   47   &    16   & 0    \\\hline 
\end{tabular}
\end{center}

\end{table*}


\begin{table*}

\caption{The information of the spectral clustering result of UMAP standardized parameters dataset embedding. There are two repeater clusters, the apparently non-repeaters in repeater clusters are classified as repeater candidates.\label{table:rate3}}
\renewcommand\arraystretch{1.5}    
\begin{center}
\begin{tabular}{c|c|c|c|c}
\hline

\multicolumn{1}{c|}{ }  &     Total number        &     Repeater number  	 &     	Repeater candidate number  & Apparently non-repeater number        \\\hline

One-off cluster 1    &   149  &   2   &   0     & 147     \\\hline
One-off cluster 2    &   15  &   0   &   0     & 15     \\\hline
One-off cluster 3    &   9  &   0   &   0     & 9     \\\hline
One-off cluster 4    &   187  &   3   &   0     & 184     \\\hline
Repeater cluster 1    &   141  &   40   &    101    & 0   \\\hline 
Repeater cluster 2    &   93  &   49   &    44   & 0    \\\hline 
\end{tabular}
\end{center}

\end{table*}

According to the results presented, the Resnet image feature data performs better among the two datasets in classifying FRBs. In addition to higher recall, classifying FRBs using image data is more model-independent.

\begin{figure*}
    \begin{center}
    \includegraphics[width=0.90\textwidth]{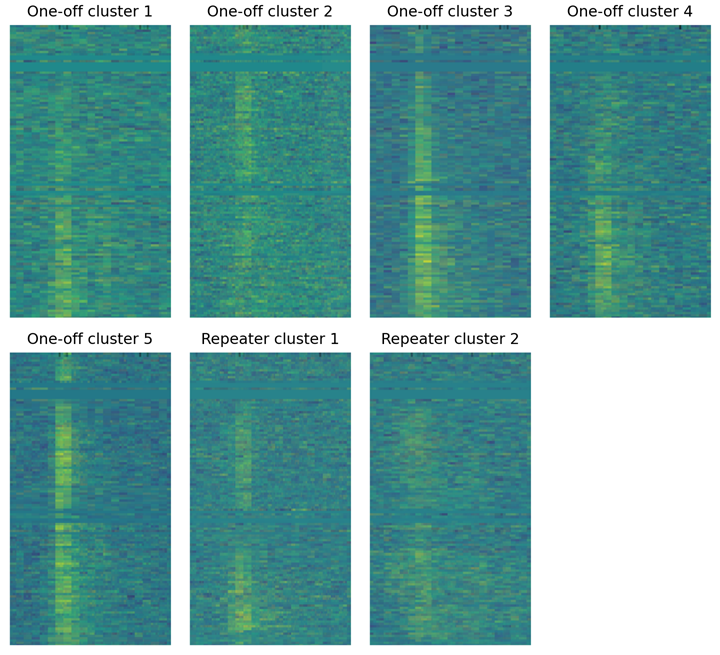}
    \caption{The representative images of each cluster for the Resnet image feature. Each figure is obtained by stacking all the members in the specific cluster. The detailed component information of each image is shown in Table~\ref{table:rate2}.}
    \label{figure:stack_res}
    \end{center}
\end{figure*}

\begin{figure*}
    \begin{center}
    \includegraphics[width=0.80\textwidth]{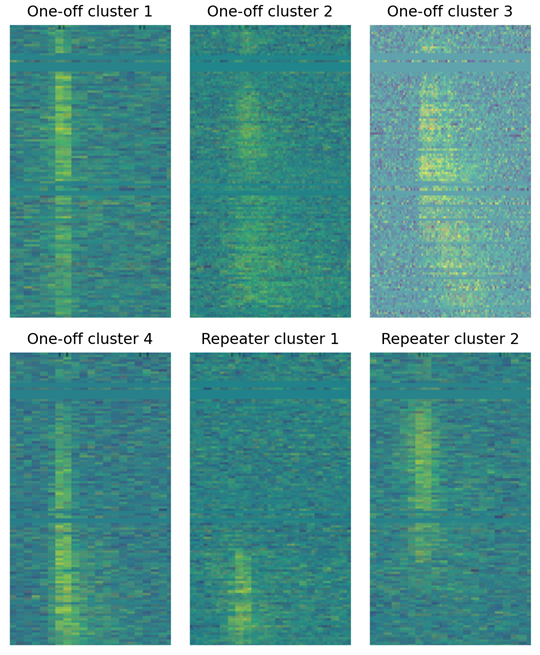}
    \caption{The representative images of each cluster for the standardized parameters. Each figure is obtained by stacking all the members in the specific cluster. The detailed component information of each image is shown in Table~\ref{table:rate3}.}
    \label{figure:stack_para}
    \end{center}
\end{figure*}

To better understand the clustering of the datasets, representative images of each cluster were presented by stacking them together. For the Resnet image feature data in Figure \ref{figure:stack_res}, it was found that narrowband FRBs of higher and lower frequencies are mixed together in repeater clusters. Additionally, one-off cluster 4, located on the left side of the gap between mixture and pure non-repeaters, is composed mostly of narrowband FRBs. In Figure \ref{figure:stack_para}, for the standardized parameters data, repeater cluster 1 represents lower frequency narrowband FRBs, while repeater cluster 2 represents higher frequency narrowband FRBs.


To investigate how different input types affect the UMAP clustering results, we compared the FRB classification outcomes of two input types in our work with one input type in~\citet{umap_frb}. Table \ref{table:input_compare} presents the detailed comparison results.
The spectral clustering of the Resnet image feature dataset and thirteen standardized parameters classified 79.0\% of apparently non-repeaters with the same label.
The spectral clustering of the Resnet image feature dataset and the HDBSCAN clustering of thirteen parameters from~\citet{umap_frb} classified 71.7\% of apparently non-repeaters with the same label.
The spectral clustering of thirteen standardized parameters and the HDBSCAN clustering of thirteen parameter parameters classified 85.8\% of apparently non-repeaters with the same label.

\begin{table}
\caption{The clustering label similarity of apparently non-repeaters from four types of input.\label{table:input_compare}}
\renewcommand\arraystretch{1.5}    
\begin{center}
\begin{tabular}{@{\extracolsep{\fill}}c|c|c|c}
\hline

\multicolumn{1}{c|}{Dataset}  & Dataset  &     Label similarity              \\\hline

Resnet Image Feature    &   Standardized Parameters  & 79.0\%    \\\hline 
Resnet Image Feature    &   Parameters  & 71.7\%    \\\hline 
Standardized Parameters    &   Parameters  & 85.8\%    \\\hline 
\end{tabular}
\end{center}
\end{table}

\subsection{Feature importance}

\begin{figure*}
    \begin{center}
    \includegraphics[width=1.00\textwidth]{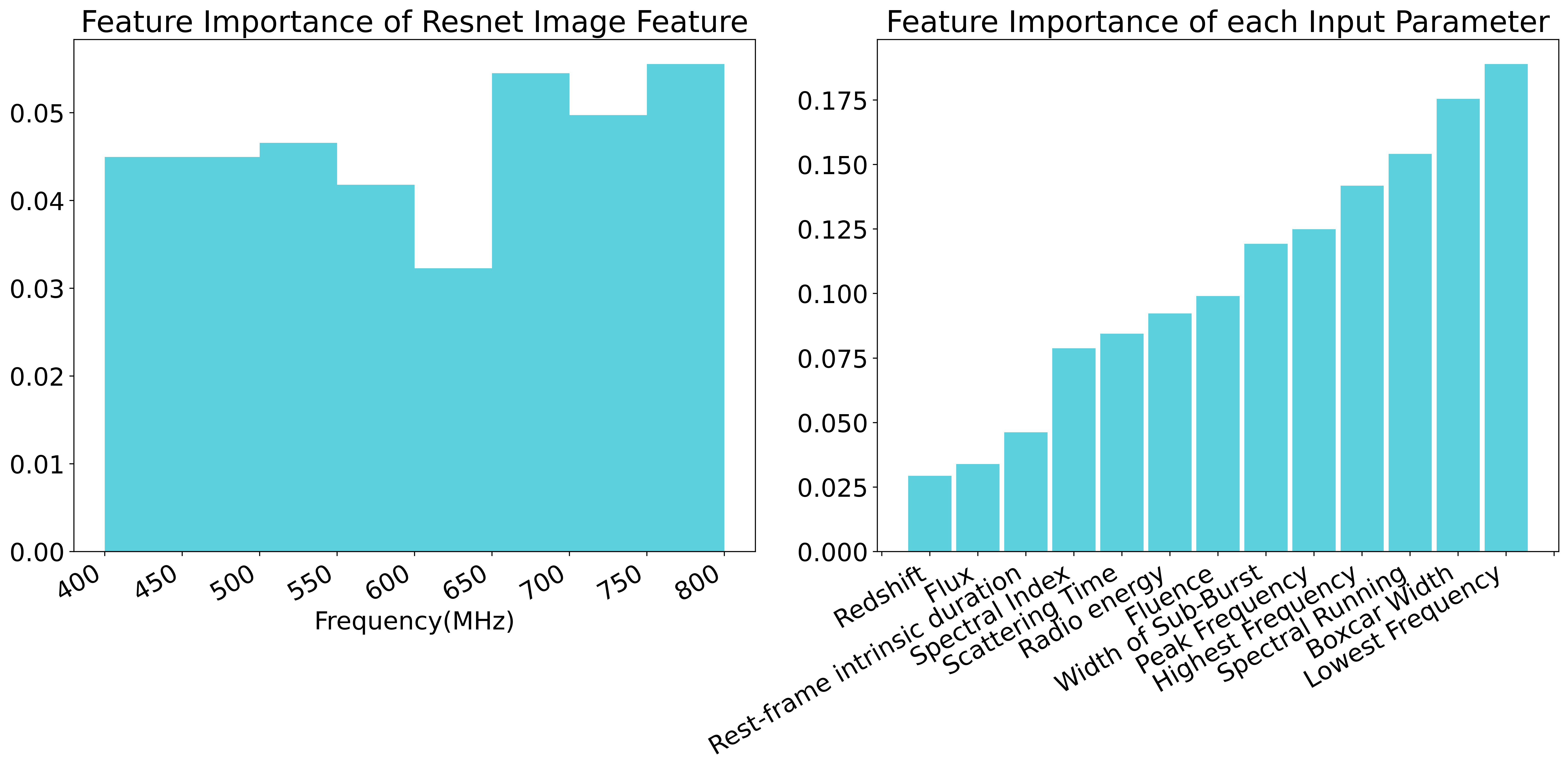}
    \caption{The left panel shows the result of permutation feature importance for the Resnet image feature data, the right panel shows the result of permutation feature importance for standardized parameters data.}
    \label{figure:6}
    \end{center}
\end{figure*}

To evaluate the contribution of each input image area or parameter, permutation feature importance was used on the UMAP models, as described in~\citep{permutation_feature}. The values of each feature were randomly permuted, and the resulting decrease in recall was used as an indicator of feature relevance. The left panel of Figure \ref{figure:6} displays the permutation feature importance for the Resnet image feature data, with a focus on the frequency channel since the time length differs for each image. However, significant fluctuation in the results was not found after permuting.
The right panel of Figure \ref{figure:6} presents the permutation feature importance of the thirteen standardized observation parameters. The lowest frequency was found to be the most important feature, influencing 18.9\% of the final result, which is only slightly more significant than the boxcar width, which impacts 17.5\%. There is no distinct boundary between the importance of different parameters, and no single parameter dominates the clustering result.

\subsection{Combination of spectrograms and physical parameters}

\begin{figure}
    \begin{center}
    \includegraphics[width=0.50\textwidth]{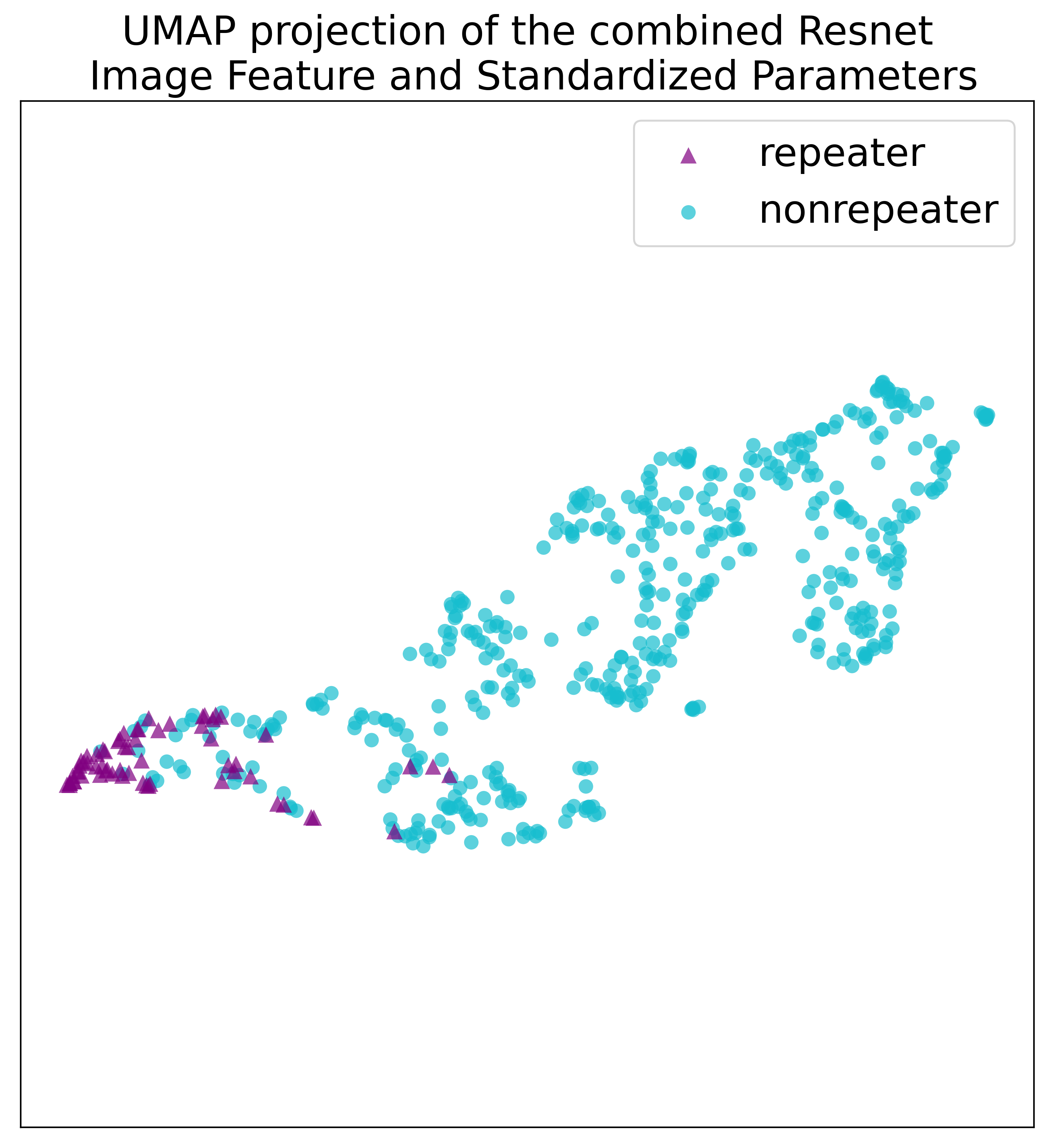}
    \caption{This figure shows the combined result of Resnet image feature and standardized parameters.}
    \label{figure:combined}
    \end{center}
\end{figure}

536 sources were available with both spectrograms and physical parameters at the same time. Additionally, we combined the array of images with standardized parameters and used it as input for UMAP. Figure \ref{figure:combined} shows the outcome of combining these two types of image data with standardized parameters data. However, there were no significant differences observed in the outputs compared to those presented in Figure \ref{figure:resimg}.

\section{Conclusion}
\label{sec:Dis}
Our findings suggest that UMAP and t-SNE, two nonlinear dimensionality reduction models, have the potential to differentiate repeaters from apparently non-repeaters in FRBs. 
We applied these models to the standardized parameter data of 594 sub-bursts and 535 FRB images data~\citep{chimecatalog}. 
Both methods using two kinds of inputs were able to differentiate repeaters from apparently non-repeaters. 
Our results do not reveal distinctly separated clusters as in~\cite{umap_frb} because we have followed the standardizing process.
In the Resnet image feature data, repeaters were tightly clustered together in a specific region,
while in the standardized parameter data, they were spread in a quarter of the two-dimensional embedding plane.
The clustering recall using our standardized parameter data is 94.7\%, which is lower than the Resnet image feature data recall of 100.0\%. 
These findings suggest that classifying FRBs using image data is a more model-independent method, with the potential to become a key method of future FRB classification.

We compared our predictions with the CHIME/FRB discovery of 25 new repeaters, and discovered that there were six repeaters in \cite{chime20repeater} that were classified as apparently non-repeaters in \cite{chimecatalog}. The detailed information of our prediction for these repeaters is listed in Table \ref{table:prediction}. We found that the standardized parameters dataset successfully predicted five out of six FRBs, while the Resnet image feature dataset predicted four out of six. It is worth noting that 20180910A appeared to be an outlier for all methods because its spectrum resembled that of a broadband FRB in the CHIME telescope, which only has a relatively narrowband receiver (400-800\,MHz).

\begin{table*}
\caption{ The prediction of two kinds of dataset comparing to the six new repeaters~\protect\citep{chime20repeater}. The $\checkmark$ symbol means the FRB is predicted as repeater candidate, the $\times$ symbol means the FRB is predicted as apparently non-repeater. \label{table:prediction}}
\renewcommand\arraystretch{1.5}    
\begin{center}
\begin{tabular}{@{\extracolsep{\fill}}c|c|c|c}
\hline

\multicolumn{1}{c|}{FRB Source}  &    Resnet Image Feature   & Standardized Parameters           \\\hline
20190609C    &      \checkmark&   \checkmark    \\\hline 
20190226B    &     $\times$ &   \checkmark    \\\hline 
20190430C    &     \checkmark&   \checkmark\\\hline 
20190110C    &     \checkmark&   \checkmark   \\\hline 
20190113A    &     \checkmark&   \checkmark   \\\hline 
20180910A   &      $\times$ &   $\times$   \\\hline 
\end{tabular}
\end{center}
\end{table*}

Processing images is a convenient way to analyse data from different instruments and at different frequency bands, such as Parkes, GBT, FAST, and SKA.  
From representative images of each cluster, we found that repeater clusters tend to be narrowband, which is consistent with our feature importance analysis. These properties imply a difference in burst morphology between repeaters and apparently non-repeaters. 
%
More FRBs detected by instruments like the Parkes ultra-wide bandwidth (704 to 4032\,MHz) receiver~\citep{Hobbs20} would be valuable for improving the accuracy of our classification methods.
Since the FRBs detected by CHIME are all within the 400-800\,MHz range, our method can only classify FRBs within this frequency range. However, when enough FRBs are detected at higher frequency bands (> 1 GHz), we can expand our classification method to cover these bands as well.  
If the repeating and non-repeating types of FRBs are essential, our classification methods should become more effective when applied to a larger and more complete FRB sample. 




\section*{Acknowledgements}
This work is partially supported by the National Key Research and Development Program of China (2022SKA0130100), the National Natural Science Foundation of China (grant No. 12041306), the international Partnership Program of Chinese Academy of Sciences for Grand Challenges (114332KYSB20210018), the ACAMAR Postdoctoral Fellow, China Postdoctoral Science Foundation (grant No. 2020M681758), and the Natural Science Foundation of Jiangsu Province (grant Nos. BK20210998). JSW acknowledges the support from the Alexander von Humboldt Foundation.

\section*{DATA AVAILABILITY STATEMENTS}
The FRB classifying results are available from \url{https://astroyx.github.io/frbdimensionreduction.github.io/}. The data used in this work is available from \url{https://www.chime-frb.ca/catalog}. The three derived parameters are calculated following the method from~\cite{umap_frb}.




\bibliographystyle{mnras}
\bibliography{mnras} 

\bsp	
\label{lastpage}
\end{document}